# Spatial Solitons and Anderson Localization

**Kestutis Staliunas**
*PTB Braunschweig, Bundesallee 100, 38116 Braunschweig, Germany*
*kestutis.staliunas@ptb.de*

**Abstract**

Stochastic (Anderson) localization is the spatial localization of the wave-function of quantum particles in random media. We show, that a corresponding phenomenon can stabilize spatial solitons in optical resonators: spatial solitons in resonators with randomly distorted mirrors are more stable than in perfect mirror resonators. We demonstrate the phenomenon numerically, by investigating solitons in lasers with saturable absorber, and analytically by deriving and analyzing coupled equations of spatially coherent and incoherent field components.

# 1. Introduction

The phenomenon of stochastic- or Anderson localization [1] is well known in quantum mechanics, and in solid state physics. Whereas the wavefunction of a single quantum particle (e.g. of an electron) is delocalized (extended) for a homogeneous potential, it can be spatially localized (confined) for a random potential. In particular the stochastic localization is believed to be responsible for the metal-insulator transitions in semiconductors with impurities [2].

Anderson localization is usually investigated in conservative systems, and is most frequently treated by solving the linear Schrödinger equation, or a spatially discrete version of the linear Schrödinger equation (so called tight binding model). Although some controversy exists, it is now generally accepted [3], that: 1) in 1- and 2-dimensional systems the Anderson localization is absolute, i.e. occurs for infinitesimal strength of the random potential, and the wave-function decays exponentially in space; 2) in 3-dimensional systems the localization is weak, i.e. occurs only for sufficiently strong random potential; 3) in 4- and more-dimensional systems the localization due to a random potential is absent.

Here we apply the idea of stochastic localization to spatial solitons in dissipative nonlinear optical resonators. There are different pictures for the formation of spatial solitons in resonators: in one interpretation the soliton is a single spot of a stripe pattern (in 1D) or of a hexagonal pattern (in 2D) embedded in the trivial zero solution, which serves as a background [4]. A bistability between the modulated state and the zero (dark) state is necessary for this interpretation. In another interpretation the soliton is a small island of the homogeneous nonzero solution embedded in the trivial zero solution [5]. The bistability between the homogeneous bright and dark states is necessary for this interpretation. In both cases dissipative solitons are stable in some parameter range (stability balloon) due to a subtle balance between the gain, saturation, self-focusing, diffraction, and resonator detuning.

The idea put forward in this article is: if a spatially random potential tends to localize the fields in linear systems, then perhaps random potentials could provide (increased) stabilization of spatial solitons in nonlinear optical systems. This could have twofold consequences: 1) if the soliton is stable for homogeneous parameters (inside the stability balloon in parameter space) then a random potential could increase its stability; 2) if the solitons are weakly unstable for a homogeneous potential (outside the stability balloon in parameter space, but close to its boundary), then a random potential could stabilize the solitons. In the first case the negative real parts of the stability exponents would become more negative, due to the randomness of the potential. In the second case the small positive real part of stability exponents would cross the zero and become negative.

We show such localization numerically on a particular example of solitons in lasers with saturable absorber [5]. The numerical integration of the corresponding equations shows a substantial increase of the stability region of the soliton due to the random potential (Chapter 3). Next we present our attempts of analytical treatment of the problem: by considering harmonic potentials instead of stochastic ones (Chapter 4), and by deriving coupled equations for spatially coherent and incoherent field components (Chapter 5). Finally the linear stability analysis of the coupled equations is performed (Chapter 6), evidencing the soliton stabilization phenomenon.

## 2. Model

The linear Schrödinger equation:

$$\frac{\partial A(\vec{r},t)}{\partial t} = i\nabla^2 A(\vec{r},t) - iV(\vec{r})A(\vec{r},t) \qquad (1)$$

or its space-discretized version, is usually used to analyze the Anderson localization in quantum mechanics and solid state physics. Here the potential $V(\vec{r})$ is a random function, stationary in time and delta-correlated in space: $\langle V(\vec{r}_1)V(\vec{r}_2)\rangle = s\delta(\vec{r}_1-\vec{r}_2)$, with the dispersion $s$.

Many nonlinear optical resonators are well described in the paraxial, and mean field approximations by nonlinear and dissipative generalizations of the Schrödinger equation:

$$\frac{\partial A(\vec{r},t)}{\partial t} = \hat{N}(A, A^*, \nabla^2 A) + i\nabla^2 A(\vec{r},t) - iV(\vec{r})A(\vec{r},t). \qquad (2)$$

The operator $\hat{N}(A, A^*, \nabla^2 A)$ corresponds to the gain, saturation, spatial frequency filtering and other possible nonlinear and dissipative effects of resonators. The second r.h.s term of (2) corresponds to the diffraction, and the last r.h.s term correspond to the mirrors of the resonator: spherical mirrors are represented by a parabolic form of potential $V(\vec{r})$, and the rough (scattering) flat mirrors correspond to the random potential.

In particular the laser with saturable absorber was investigated, a system supporting spatial solitons [5], where the dissipative part of (2) is given by:

$$\hat{N}(A, A^*, \nabla^2 A) = \left(\frac{D_0}{1+|A|^2} - 1 - \frac{a_0}{1+|A|^2/I_a} - g\nabla^4\right)A \qquad (3)$$

i.e. consisting of saturating gain, of linear losses, of saturating losses, and of spatial filtering respectively. ($D_0$ is gain, $a_0$ is unsaturated absorption, $I_a$ is saturation intensity of the absorber; See e.g. [5] for detailed description of the model (2), (3).)

The similarity between (1) and (2) is not unexpected, since the photons are quantum particles. Therefore the stochastic localization as shown for (1) is plausible also in nonlinear optical systems.

## 3. Numerics

A numerical analysis of (2), (3) was performed in order to investigate straightforwardly the role of the randomness on the stability of spatial solitons in the 2D case. A normal split-step technique was used, imposing periodic boundary conditions. The potential is given by uncorrelated Gaussian distributed random number on every grid site. Fig.1 shows the stability range of a spatial soliton depending on the strength of the random potential. Evident is an increase of the soliton stability area due to the random potential. The spatial distributions of the intensity of the light for random and for homogeneous potentials are given in the insets in Fig.1. Evident is the decrease of the overall size of solitons, and the shift of the solitons to some favored places depending on the particular realization of the random noise. I.e. the translational invariance of the space is broken.

Fig.1 evidences the following effects of the random potential:

1) the soliton existence range shifts towards larger values of the pump $D_0$ with increasing strength of the random potential $s$. A simple interpretation is that the random potential induces effective losses (the radiation is permanently scattered into higher spatial components), and larger values of pump are required to compensate the increasing losses. The induced losses depend seemingly linearly on the strength of the random potential for larger $s$, but follow a square root law for small values of $s$. (As shown below the noise induced losses are proportional to the square root of $s$ in the limit of small amplitude of the noise.)

2) The soliton stability range increases with increasing strength of the random potential. The dependence again seems linear for larger $s$, but follows a square root law for small values of $s$.

More detailed analysis is required to determine qualitatively the increase of the stability ranges of solitons. As Fig.1. indicates the different realizations of the random potential result in a slightly different stability ranges. A time consuming averaging over many realizations of potential is then required. An alternative approach, used below, is the substitution of the random potential by a harmonic one (Chapter 4).

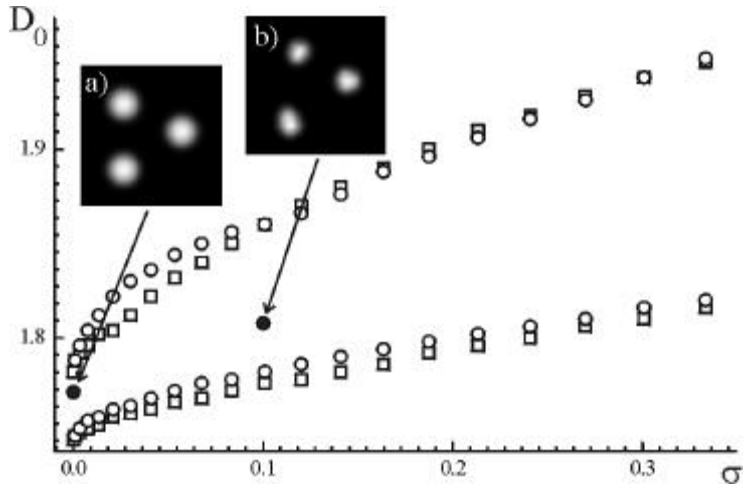

*Fig.1. Soliton stability range depending on the strength of the random potential $s$, for two realizations of the random potential, for laser with saturable absorber in 2 spatial dimensions, as obtained by numerical integration of (2), (3), with $I_a = 0.01$, $a = 5$, $g = 1$. The solitons are stable for pump values between those indicated by open circles and by squares for different realizations of the random potential. The inset a) shows the stable spatial solitons for homogeneous potential ($s = 0$) for $D_0=1.77$, and the inset b) shows the stable solitons for random potential ($s = 0.1$) for $D_0=1.8$.*

### 4. Harmonic potential

The soliton stabilization by periodic spatial perturbation is investigated for the case of one spatial dimension: $A \equiv A(x,t)$. The model (2) then reads:

$$\frac{\partial A}{\partial t} = \hat{N}(A, A^*, \frac{\partial^2 A}{\partial x^2}) + i\frac{\partial^2 A}{\partial x^2} - im(e^{ikx} + e^{-ikx})A . \qquad (4)$$

A numerical analysis of (4) with (3) was performed. Fig. 2 shows the dependence of the stability range of solitons on the strength of the spatial modulation $m$ for different wavenumbers of spatial perturbation $k$.

In contrast to expectations, the use of spatially harmonic potentials does not allow to determine more precisely the soliton stability range, compared with the case of a random potential (Fig.1). Moreover the stability range increases in general nonmonotonically with increasing amplitude of harmonic spatial modulation. This nonmonotonic dependence is related with a small number of modulation periods within the soliton width (see Fig.3.a). Evidently solitons of a particular width are favored, i.e. the fronts forming the solitons lock at distances of integer numbers of modulation periods. (The soliton stability is related with the mutual locking of fronts [6].) The width of the soliton, on the other hand depends on the amplitude of the modulation. This explains the nonmonotonic behavior of the soliton stability range. This explains also the numerical observation, that the nonmonotonic behavior is more prominent for small wavenumbers of perturbation $k$, and disappears for increasing perturbation wavenumbers.

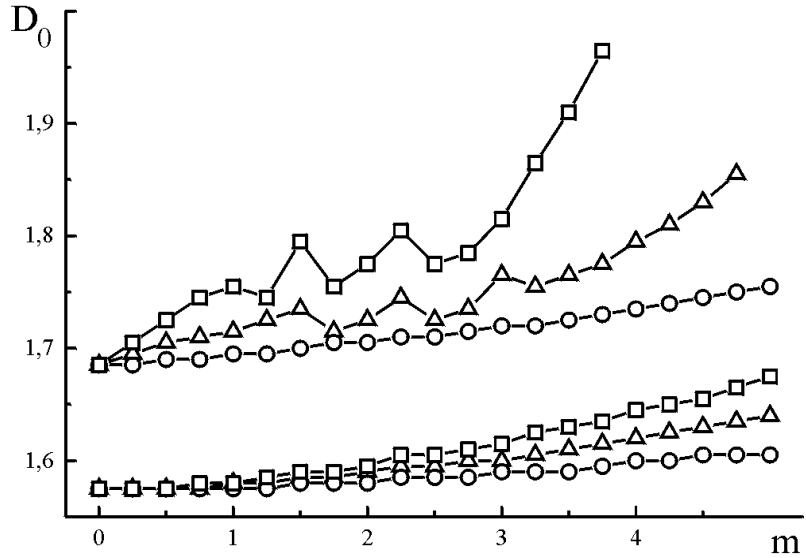

*Fig.2. Soliton stability range depending on the amplitude of harmonic spatial modulation m, for three different modulation wavenumbers k, for laser with saturable absorber in 1 spatial dimension, as obtained by numerical integration of (4), (3), with $I_a = 0.01$, $a = 5$, $g = 1$. Solitons are stable for pump values between those indicated by open circles (k = 6) by triangles (k = 5), and by squares (k = 4).*

Since the dispersion of the modulated potential corresponds to $m^2$, the likely linear dependence of the stability range for weak modulation in Fig.2 is compatible with the square root dependence in Fig.1.

The power spectrum of soliton field (Fig.3.b) consists of a central peak, corresponding to the homogeneous part of the soliton, of first order sidebands, corresponding to harmonically modulated soliton components, and of weak higher order sidebands. The sidebands are well separated, which suggests, that the soliton field can be decomposed into homogeneous and harmonic components.

$$A(x,t) = A_0(x,t) + A_{+k}(x,t)e^{ikx} + A_{-k}(x,t)e^{-ikx} \quad (5)$$

where the second and higher order sidebands are neglected.

Inserting (5) into (4) and collecting terms at different harmonics one obtains a set of coupled differential equations:

$$\frac{\partial A_0}{\partial t} = N_0(A_0, A_{\pm k}) + i\frac{\partial^2 A_0}{\partial x^2} - im(A_{+k} + A_{-k}) \quad (6.a)$$

$$\frac{\partial A_{\pm k}}{\partial t} = N_{\pm k}(A_0, A_{\pm k}) + i\left(\frac{\partial}{\partial x} \pm ik\right)^2 A_{\pm k} - imA_0 \quad (6.b)$$

where $N_i(A_0, A_{\pm k})$ are nonlinearities which couple the homogeneous and harmonic components of the field.

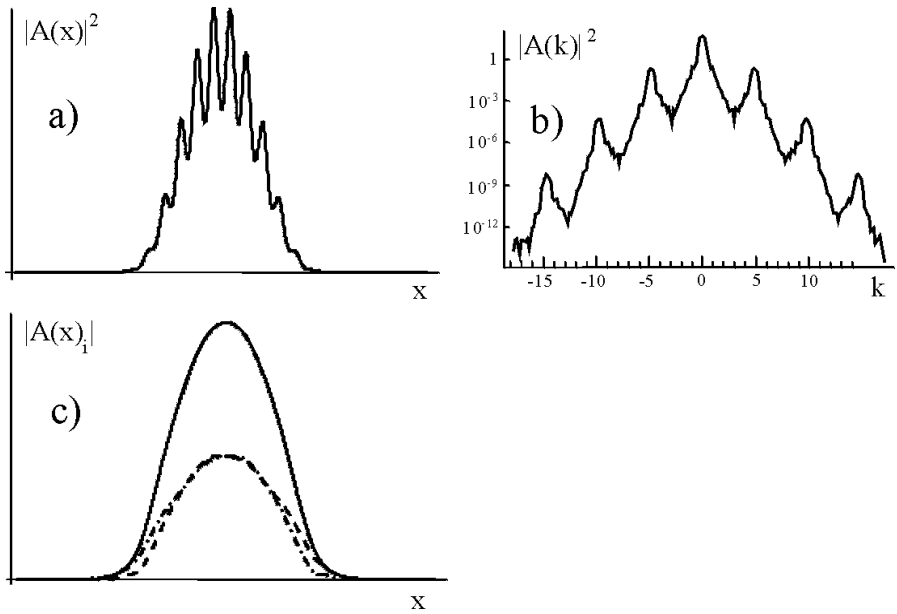

*Fig. 3. a) Field intensity of the soliton for harmonically modulated background; b) spatial power spectrum in logarithmic representation; as obtained by numerical integration of (4), (3) with: $I_a = 0.01$, $a = 5$, $g = 1$, $k = 5$, $m = 5$. c) homogeneous (solid line) and harmonic (dashed, and point dashed lines) field components, corresponding to central and first order sidebands of spectra given in b). The envelope of homogeneous field component in c) is arbitrarily scaled in order to enhance visibility.*

The system has no simple solution even for the simplest form of nonlinearity, e.g. for the nonlinearity following from the Complex Ginzburg-Landau equation: $\hat{N}(A, A^*) = (p - |A|^2)A$.

However, the decomposition (5) allows to gain insight into the role of the harmonic modulation on the stability of solitons.

In particular the numerically obtained spatial Fourier spectrum was decomposed into components centered about the central and first sidebands, then, using inverse Fourier transform, the homogeneous and harmonic field components $A_0(x,t)$ and $A_{\pm k}(x,t)$ were reconstructed. The profiles of the decomposed field are shown in Fig.3.c. The following conclusions can be made:

1) The profiles of harmonic field components follow approximately the profile of the homogeneous component. The left- and right scattered fields are, however, shifted left and right respectively, in accordance with the advection terms in (6.b). The amplitudes of the harmonic field components are weaker than the amplitude of the homogeneous field component. In particular the following scaling holds: $|A_{\pm k}|/|A_0| \propto m/k^2$, in the limit of $m \ll 1$ and $k \gg 1$. This asymptotic scaling is valid for general nonlinearities, the single assumption being the stability of the solution. Our numerical analysis confirms this asymptotic scaling (note that the amplitude of the homogeneous field component in Fig.3 was scaled in order to enhance visibility).

2) The profile of the total scattered field $A_+(x,t) = A_{+k}(x,t) + A_{-k}(x,t)$ is very similar to the profile of the homogeneous component $A_0(x,t)$. The profile of the total scattered field of the soliton is only slightly broader (diffused) than that of the homogeneous field of soliton. This is compatible with the fact that right and left scattered components are respectively advected. This results in a broadening of the envelope of the overall scattered field $A_+(x,t)$.

3) The difference of the left- and right scattered fields $A_-(x,t) = A_{+k}(x,t) - A_{-k}(x,t)$ is much smaller than the total scattered field: $|A_-| \ll |A_+|$. As shown below the scaling $|A_-|/|A_+| \propto 1/k$ holds in the limit of $k \gg 1$.

The above observations allow to consider the following hierarchy of smallness of the fields: $|A_-| \ll |A_+| \ll |A_0|$, and suggest to rewrite the system (6) in terms of $A_-(x,t)$, $A_+(x,t)$ and $A_0(x,t)$:

$$\frac{\partial A_0}{\partial t} = N_0(A_0, A_\pm) + i\frac{\partial^2 A_0}{\partial x^2} - imA_+ \tag{7.a}$$

$$\frac{\partial A_+}{\partial t} = N_+(A_0, A_\pm) + i\left(\frac{\partial^2}{\partial x^2} - k^2\right)A_+ - 2k\frac{\partial}{\partial x}A_- - 2imA_0 \tag{7.b}$$

$$\frac{\partial A_-}{\partial t} = N_-(A_0, A_\pm) + i\left(\frac{\partial^2}{\partial x^2} - k^2\right)A_- - 2k\frac{\partial}{\partial x}A_+ \tag{7.c}$$

Assuming the above hierarchy of fields, $A_-$ can be adiabatically eliminated from (7.c) leading to the formal expression:

$$A_- = \hat{I}_-\left(A_0, k^2, \frac{\partial^2}{\partial x^2}\right)^{-1} 2k\frac{\partial}{\partial x}A_+. \tag{8}$$

Here $\hat{I}_-(A_0, k^2, \partial^2/\partial x^2)$ is the operator associated with the most unstable eigenmode (eigenvalue with the largest real part) of a stationary solution of the autonomous part of (7.c).

$\hat{I}_-(A_0, k^2, \partial^2/\partial x^2)$ is in general an operator, however in the limit $k \gg 1$, the spatial derivative can be neglected, and $I_-(k)$ is an algebraic function of $k$:

$$A_- = \frac{2i}{k}\frac{\partial}{\partial x}A_+ + O(k^{-3})\frac{\partial}{\partial x}A_+ + O(k^{-3})\frac{\partial}{\partial x}A_+^* \tag{9}$$

Inserting (9) into (7.b) one obtains:

$$\frac{\partial A_+}{\partial t} = N_+(A_0, A_\pm) - ik^2 A_+ - 3i\frac{\partial^2 A_+}{\partial x^2} - 2imA_0 \tag{10}$$

which indicates that the net diffraction coefficient for the scattered field has a negative sign and is three times larger than that of the homogeneous component of the solution.

The system of coupled equations for homogeneous field component (7.a) and for the harmonic field components (10) gives an interpretation for the stabilization of spatial solitons. The system (7.a), (10) can be identified with the Turing equation system for local activator and lateral inhibitor [7]. A more recent analysis show that a diffusion of the inhibitor field (in particular the diffusion of population inversion field of a laser in [8]), as well as a diffraction of the inhibitor field (in particular the diffraction of the pump field of optical parametric oscillators in [9]) leads to the enhancement of the modulational instability, and consequently, to stabilization of spatial solitons. In [8] and [9] the original Turing pattern formation mechanism was generalized to diffracting fields, and in general to nonlocal fields, where nonlocality can be caused by diffusion as well as by diffraction, or both. (10) shows that one encounters here the generalized Turing pattern formation mechanism, where the net diffraction of harmonic field components plays the role of a nonlocality.

Finally, the stationary solution of (10) in the above limits of smallness is:

$$A_+(x) = -\frac{2m}{k^2}\left(1 + \frac{3}{k^2}\frac{\partial^2}{\partial x^2}\right)A_0(x), \tag{11}$$

which means that the harmonic field component is proportional to the homogeneous field component (first term), but slightly diffused (second term). In particular the widths of harmonic and homogeneous parts of the soliton $x_1$ and $x_0$ are related by: $x_1 = x_0\sqrt{1 + 6/(kx_0)^2}$. Our numerical results are compatible with (11), i.e. we observe a broadening of harmonic field components relative to homogeneous field component. The broadening is independent on $m$ and weakly dependent on $k$, in accordance with (11).

## 5. Coherent - incoherent field decomposition

The results obtained for harmonic potentials can be generalized to a random potential. However, restrictions, as described below, must be imposed on the random potential in order to exclude the effects of so called "naive stochastic localization".

It the potential is completely random, then there exist local minima of the potential. Evidently the soliton moves towards one of the local minima, and remains pinned and stabilized by it. Depending on the realization of the random potential the minima can be sufficiently deep and broad, and then the soliton is no more a soliton in a conventional sense (as stabilized by nonlocal and nonlinear effects), but just a lowest eigenmode of the potential well. The potential around its minimum can be approximated by a parabola. For a sufficiently deep potential well the stability range of the soliton increases up to the bistability range of the lowest eigenmode of the parabolic potential well. The latter is comparable with the bistability range of plane waves of (2), (3).

The above naive localization is due to the variation of the random potential on a large spatial scale. In order to exclude the effects of naive localization one must explicitly separate spatial scales: to introduce a large spatial scale $X$ related with the envelope of the soliton, and the small spatial scale $x$, related with the stochastic potential. The assumption that the stochastic potential does not depend on the large spatial scale imposes that the potential can be expanded as $V(x) = \sum_{n \neq 0} m_n \exp(ik_n x)$, where $k_n = k_0 n$ and $k_0 = 2\pi/x_0$ depends on the definition of spatial scales: $x < x_0 < X$. The above Fourier expansion means that the random potential is periodic with the period $x_0$, which however is a necessary condition for its stationarity on the large spatial scale $X$. The optical field is expanded as:

$$A(x,t) = A_0(X,t) + \sum_{n \neq 0} A_n(X,t) \exp(ik_n x) \tag{12}$$

where $A_0(X,t)$ is the spatially coherent component, and $A_n(X,t), n \neq 0$ are harmonic components of the field, both depending on the large spatial scale coordinate only. Upon introducing the envelope of the spatially incoherent component of the field, $A_{inc}(x,t) = \sum_{n \neq 0} A_n(X,t) \cdot m_n^*$, the temporal evolution of the coherent field component $A_0(X,t)$ (the analog of equation (7.a)) is:

$$\frac{\partial A_0}{\partial t} = N_0(A_0, A_{inc}) + i\frac{\partial^2 A_0}{\partial x^2} - iA_{inc} \tag{13.a}$$

and the temporal evolution of $A_{inc}(X,t)$, as obtained by the use of (10) is:

$$\frac{\partial A_{inc}}{\partial t} = N_{inc}(A_0, A_{inc}) - i\sum_{n \neq 0} k_n^2 A_n(X,t) \cdot m_n^* - 3i\frac{\partial^2 A_{inc}}{\partial x^2} - 2i\sigma A_0. \tag{13.b}$$

here $\sigma = \sum |m_n|^2$ is the dispersion of the random potential.

The equation system (13) is not closed due to the second term in (13.b). The closure is, however, possible noticing that the scattered field components are proportional to the total scattered field, within the limit of smallness used: $k = O(\varepsilon^{-1})$ (see discussion on enslaving of the fields in Chapter 4), Then one can substitute $i\sum_{n \neq 0} k_n^2 A_n(X,t) \cdot m_n^*$ by $ik_{av}^2 A_{inc}(X,t)$, where $k_{av}^2 = \sum_{n \neq 0} |m_n|^2 / \sum_{n \neq 0} k_n^{-2} |m_n|^2$ has a meaning of average wavenumber of the spatially incoherent field. Then (13) simplifies to:

$$\frac{\partial A_0}{\partial t} = N_0(A_0, A_{inc}) + i\frac{\partial^2 A_0}{\partial x^2} - iA_{inc} \tag{14.a}$$

$$\frac{\partial A_{inc}}{\partial t} = N_{inc}(A_0, A_{inc}) - ik_{av}^2 A_{inc} - 3i\frac{\partial^2 A_{inc}}{\partial x^2} - 2is A_0 \tag{14.b}$$

The system (14) describes the dynamics of the coupled coherent and incoherent parts of the radiation. Its form is completely analogous to the equation system derived for harmonic potential (7.a), (10). The system (14) is straightforwardly extendable to 2D case (by substituting spatial derivatives by Laplace operators). However the extension to 3D case is problematic, due to divergences by calculating $k_{av}$. The latter is compatible with the result from the usual Anderson localization in conservative system, stating that the localization is absolute in 1D and 2D cases only.

## 6. Linear stability analysis

The enhanced diffraction of the inhibitor leads in general to a reduction of the stability of the homogeneous solution, and correspondingly to enhancement of the stability of solitons [8], [9]. We show this decrease of stability of the homogeneous solution by performing the linear stability analysis of the Complex Ginzburg Landau equation. We use the Complex Ginzburg Landau nonlinearity for its simplicity, but also for universality of the results. As long as the modulation of the homogeneous solution is small $m \ll 1$, the results for Complex Ginzburg Landau equation are also asymptotically valid for arbitrary soliton supporting systems, e.g. a laser with saturable absorber.

The separation into spatially coherent and incoherent field components for the Complex Ginzburg-Landau equation, as following from (7a) (10), (also from (14)) leads to:

$$\frac{\partial A_0}{\partial t} = pA_0 - |A_0|^2 A_0 + i\frac{\partial^2 A_0}{\partial x^2} - imA_1 \tag{15.a}$$

$$\frac{\partial A_1}{\partial t} = pA_1 - 2|A_0|^2 A_1 - A_0^2 A_1^* - ik^2 A_1 - 3i\frac{\partial^2 A_1}{\partial x^2} - 2imA_0 \tag{15.b}$$

The homogeneous and stationary solution of (15) does not result in simple algebraic expression. The asymptotics of stationary solutions (for $k \gg 1$, and $m \ll 1$) are given by:

$$r_0 = 1; \qquad r_1 = 2m/k^2;$$

$$\varphi = 2/k^2 - p; \qquad \omega = 2m^2/k^2; \tag{16}$$

Here $A_{j,0} = r_j \exp(i\varphi_j + i\omega t)$, and $\varphi = \varphi_1 - \varphi_0$ is the phase shift between the scattered and coherent field components. The parameters of the stationary solution, as obtained by numerical solution of (15) is given in Fig.4.

The linear stability analysis of the stationary solution: $A_j = A_{j,0} + a_j \exp(\lambda(q)t + iqx)$ involves a diagonalization of the (4*4) matrix, and was performed only numerically. A typical result is given in Fig.5.

The general tendency as following from the analysis, is that the increase of coupling $m$ between the field components reduces the stability of homogeneous solution. Although the homogeneous solution remains always stable for the Complex Ginzburg-Landau equation in the modulationally stable regime (the real part of Lyapunov exponent do not become positive), however, the "soft modes" become less damped: the real part of Lyapunov exponent approaches zero from below).

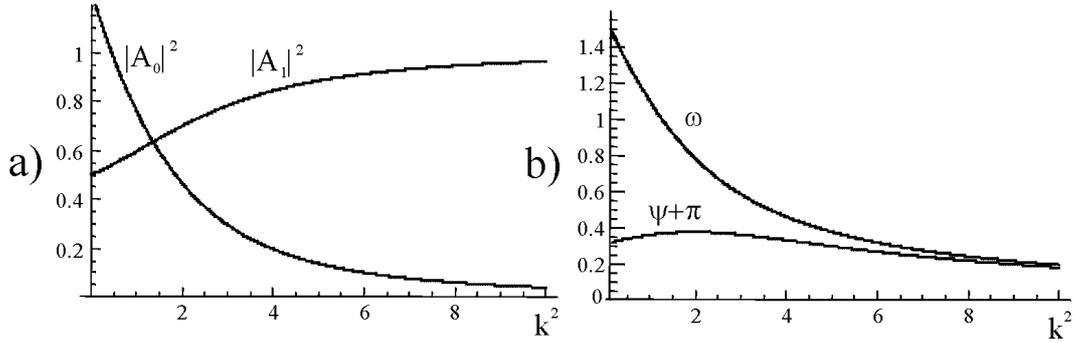

*Fig.4.a) Intensities of spatially coherent and incoherent part of the field, and b) the frequency of the field, and the phase difference between scattered and homogeneous field components, as obtained from numerical integration of (15). Parameters: $p = 1$, $m = 1$*

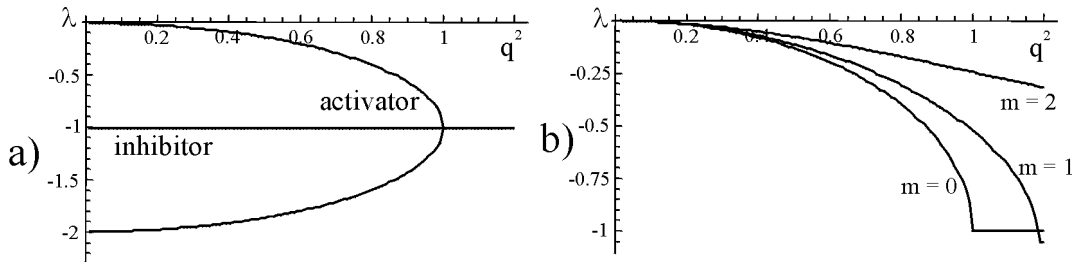

*Fig.5.a) Real part of the Lyapunov exponents as obtained by numeric linear stability analysis of stationary solution of (15) in absence of coupling $m = 0$. b) Largest real parts of the Lyapunov exponents for different values of coupling. Other parameters: $p = 1$, $k^2 = 10$.*

## 7. Conclusions

The numerical and analytical investigations lead to a somewhat counterintuitive result, namely that spatial solitons in resonators with randomly distorted (randomly scratched) mirrors and randomly scattering optical elements should be more stable than the solitons in resonators with perfect mirrors and perfect optical elements.

We show the phenomenon numerically for a particular system, a laser with saturable absorber. However, the separation into spatially coherent and incoherent components, and the derivation of coupled equations (7a) (10) and (14) generalize the phenomenon. The enhancement of the net diffraction of spatially incoherent component is responsible for the stabilization of spatial solitons, as following from the Turing pattern formation mechanism [7] and from its generalizations [8] and [9].

We note finally, that fully incoherent solitons as recently shown experimentally [10], and analyzed theoretically [11,12], where a beam of incoherent light is self-trapping due to a nonlinearity. The solitons reported here are, however, more similar to the coherent ones [4,5]: only a small incoherent part of the radiation appears due to a scattering of coherent light on the random potential. And the stabilization of the soliton occurs due to the perturbative influence of the small spatially incoherent part of the soliton on the dominating coherent part.

**Acknowledgement**
The work has been supported by Sonderforschungsbereich 407 of Deutsche Forschungsgemeinschaft, by Network PHASE of European Science Foundation, and by Centrum of Excellence in Vilnius University (CEBIOLA). Discussions with C. O. Weiss, S. Longhi, and G. Valiulis are gratefully acknowledged.